\documentclass[12pt]{article}
\usepackage{graphicx,epsfig, color}
\usepackage{amssymb,amsmath,amscd,amsthm}

\usepackage{graphicx}
\usepackage[active]{srcltx}

\newtheorem{theorem}{Theorem}[section]

\newtheorem{lemma}[theorem]{Lemma}

\date{}

\begin{document}
 \author{ Evgeny Lakshtanov\thanks{Department of Mathematics, Aveiro University, Aveiro 3810, Portugal.  This work was supported by Portuguese funds through the CIDMA - Center for Research and Development in Mathematics and Applications and the Portuguese Foundation for Science and Technology (``FCT--Fund\c{c}\~{a}o para a Ci\^{e}ncia e a Tecnologia''), within project UID/MAT/0416/2013 (lakshtanov@ua.pt).} \and
 Jorge Tejero\thanks{Universidad Aut\'onoma de Madrid and ICMAT. The work was partially supported by the ERC grants ERC-301179 and ERC-277778 and the MINECO grants SEV-2015-0554 and MTM-2013-41780-P (Spain) (jorge.tejero@icmat.es).}  \and
 Boris Vainberg\thanks{Department
of Mathematics and Statistics, University of North Carolina,
Charlotte, NC 28223, USA. The work was partially supported  by the NSF grant DMS-1410547 (brvainbe@uncc.edu).}}

\title{Uniqueness in the inverse conductivity problem for complex-valued Lipschitz conductivities in the plane}

\maketitle

\begin{abstract}
We consider the inverse impedance tomography problem in the plane.
Using Bukhgeim's scattering data for the Dirac problem,
we prove that the conductivity is uniquely determined by
the Dirichlet-to-Neuman map.
\end{abstract}

\textbf{Key words:}
Bukhgeim's scattering problem, inverse Dirac problem, inverse conductivity problem, complex conductivity.

\section{Introduction}

Let $\mathcal O$ be a bounded domain in $\mathbb R^2$.
The electrical impedance tomography problem (e.g., \cite{borcea}) concerns determining the impedance in the interior of $\mathcal O$, given simultaneous
measurements of direct or alternating electric currents and voltages at the boundary $\partial \mathcal O$.
If the magnetic permeability can be neglected, then the problem can be reduced to the inverse conductivity problem (ICP), i.e., to the problem  of reconstructing function $\gamma(z), z=(x,y) \in \mathcal O$, from the set of data $(u|_{\partial \mathcal O},\gamma\frac{\partial u}{\partial\nu}|_{\partial \mathcal O})$, dense in an adequate topology, where
\begin{equation}\label{set27A}
\mbox{div}(\gamma \nabla u(z)) =0, ~  z\in \mathcal O.
\end{equation}
 Here $\nu$ is the unit outward normal to $\partial \mathcal O$, $\gamma(z) = \sigma(z)+ i \omega\epsilon(z)$, where $\sigma$ is the electric conductivity and $\epsilon$ is the electric permittivity. If the frequency $\omega$ is negligibly  small, then one can assume that $\gamma$ is a real-valued function, otherwise it is supposed  to be a complex-valued function.

An extensive list of references on the tomography problem can be found in the review \cite{borcea}. Here we will mention only the papers that seem to be particularly related to the present work.

For real $\gamma$, the inverse conductivity problem has been reduced to the inverse problem for the Schr\"{o}dinger equation. The latter was solved by Nachman in \cite{nachman} in the class of twice differentiable conductivities. Later, Brown and Uhlmann \cite{bu} reduced the ICP to the inverse problem for the Dirac equation, which has been solved in  \cite{bc1}, \cite{sung1}. This approach requires the existence of only one derivative of $\gamma$. The authors of \cite{bu} proved the uniqueness for the ICP. Later, Knudsen and Tamasan \cite {knud} extended this approach and obtained a method to reconstruct the conductivity. Finally, the ICP has been solved by Astala and Paivarinta in \cite{ap} for real conductivities when both $\gamma-1$ and $1/\gamma-1$ are in $L^\infty_{\rm comp}(\mathbb R^2)$.

If a complex conductivity has at least two derivatives, then one can reduce equation (\ref{set27A}) to the Schr\"{o}dinger equation
and apply the method of Bukhgeim \cite{bukh} (or some of the works extending this method, such as \cite{BIY15}, \cite{lnv} or \cite{T}).
This approach does not work in the case of only one time differentiable complex valued conductivities.
On the other hand, the work of Francini \cite{fr}, where the ideas of \cite{bu} were extended to deal with
complex conductivities with small imaginary part,
are not applicable to general complex conductivities due to possible existence of the so called {\it exceptional points}.
In \cite{lbcond}, Lakstanov and Vainberg extended the ideas of \cite{lnv} to apply the $\overline \partial$-method
in the presence of exceptional points and reconstructed generic conductivities
under the assumption that $ \gamma-1 \in W^{1,p}_{\rm{comp}}(\mathbb R^2), p>4, $ and $\mathcal F(\nabla \gamma )\in L^{2-\varepsilon}(\mathbb R^2)$ (here $\mathcal F$ is the Fourier transform).

In this paper, we will prove that complex-valued Lipschitz conductivities are uniquely determined by information on the boundary.
Since we use the standard reduction of (\ref{set27A}) to the Dirac equation followed by the solution of the inverse problem for the Dirac equation, the condition on $\gamma$ can be restated in the form $Q \in L^{\infty}_{comp}(\mathbb R^2) $, where $Q$ is the potential in the Dirac equation. Our present result is based on a development of the Bukhgeim approach, combined with some of the arguments of Brown and Uhlmann from \cite{bu}. The statement of our main theorem is the following.

\begin{theorem}\label{uniqueness} Let $\mathcal O$ be a bounded Lipschitz domain in the plane and let $\gamma_1, \gamma_2$ be complex-valued Lipschitz conductivities. Then
$$ \Lambda_{\gamma_1} = \Lambda_{\gamma_2} \, \Rightarrow \, \gamma_1 = \gamma_2, $$
where $\Lambda_{\gamma_j}$ is the Dirichlet-to-Neumann map for the conductivity $\gamma_j$.
\end{theorem}

The Dirichlet-to-Neumann (DtN) map $\Lambda_\gamma : H^{1/2}(\mathcal{\partial O}) \to H^{-1/2}(\mathcal{\partial O})$ is defined by
$$\Lambda_\gamma [u\vert_{\partial \mathcal{O}}] = \gamma \frac{\partial u}{\partial \nu}\vert_{\partial \mathcal{O}},$$
where $u$ is a solution to \eqref{set27A} and $\frac{\partial u}{\partial \nu}$ is the normal derivative of $u$ at the boundary of $\mathcal{O}$.
Function $\gamma \frac{\partial u}{\partial \nu}\in H^{-1/2}(\mathcal{\partial O})$ is defined as such an element of the space dual to $H^{1/2}(\mathcal{\partial O})$ that
$$\langle \gamma \frac{\partial u}{\partial \nu}, v \rangle = \int_{\mathcal{ O}} \gamma \nabla u \cdot \nabla vdxdy$$
for each $v \in H^{1}(\mathcal{O})$.

In section \ref{outline}, we will describe our approach, stating the most relevant results.
All the proofs will be given in section \ref{proofs}.

\section{Main steps}\label{outline}

\subsection{Reduction to the Dirac equation}

From now on, we will consider $z$ as a point of a complex plane:  $z=x+iy\in\mathbb C$, and $\mathcal O$ will be considered as a domain in $\mathbb C$.
The following observation made in \cite{bu} plays an important role. Let $u$ be a solution of (\ref{set27A}) and let $\partial = \frac 1 2 \left (\frac{\partial}{\partial x} - i\frac{\partial}{\partial y} \right )$. Then the pair
$\phi=\gamma^{1/2}(\partial u, \overline{\partial} u)^t$
satisfies the Dirac equation
\begin{equation}\label{firbc}
\left ( \begin{array}{cc} \overline{\partial} & 0 \\ 0  & \partial \end{array} \right ) \phi = q {\phi}, \quad   z\in \mathcal O,
\end{equation}
where
\begin{eqnarray}\label{char1bc}
q(z)=\left ( \begin{array}{cc}0 &q_{12}(z) \\
q_{21}(z) &  0\end{array} \right ), \quad q_{12}=-\frac{1}{2}\partial \log \gamma, \quad q_{21}=-\frac{1}{2}\overline{\partial }\log \gamma.
\end{eqnarray}
 Thus the inverse Dirac scattering problem is closely related to the ICP.
If $q$ is found and the conductivity $\gamma $ is known at one point $z_0\in \overline{\mathcal O}$, then $\gamma$ in $\mathcal O$ can be immediately found from (\ref{char1bc}).

From now on, we will use a different form of equation (\ref{firbc}): instead of Beals-Coifmann notations $\phi=(\phi_1,\phi_2)^t$, we will rewrite the equation in Sung notations: $\psi_1=\phi_1,\psi_2=\overline{\phi_2}$. We will consider the equation in the whole plane by extending the potential $q$ outside  $\mathcal O$ by zero. Then the {\it vector} $\psi=(\psi_1,\psi_2)^t$ is a solution of the following system
\begin{equation}\label{fir}
\overline{\partial }\psi = Q \overline{\psi}, \quad z\in \mathbb C,
\end{equation}
where
\begin{eqnarray}\label{char1}
Q(z)=\left ( \begin{array}{cc}0 &Q_{12}(z) \\
Q_{21}(z) &  0\end{array} \right ), \quad Q_{12} = q_{12}, \quad Q_{21} = \overline{q_{21}}.
\end{eqnarray}

\subsection{Solving the Dirac equation for large $|\lambda|$}
 Let $\psi$ be a {\it matrix} solution of (\ref{fir}) that depends on parameter $\lambda \in \mathbb C$ and has the following behavior at infinity
\begin{equation}\label{lim}
\psi(z,w,\lambda) e^{-\lambda(z-w)^2/4} \rightarrow I, ~ z \rightarrow \infty.
\end{equation}
Note that the unperturbed wave
\begin{equation}\label{exp}
\varphi_0(z,\lambda,w):=e^{\lambda (z-w)^2/4}, \quad w,\lambda\in \mathbb C,
\end{equation}
 depends on the spacial parameter $w$ and the spectral parameter $\lambda$,
and  grows at infinity exponentially in some directions. The same is true for the elements of the matrix $\psi(z,\lambda,w)$.  Let us stress that, contrary to the standard practice, we consider function $\psi$ (and other functions defined by $\psi$) for all complex values of $\lambda$, not just for $i\lambda,~\lambda>0$. This allows us to generalize the Bukhgeim method to the case of potentials in $L^\infty_{com}(\mathbb R^2)$. From the technical point of view, this allows us to use the Hausdorff-Young inequality.

Problem (\ref{fir})-(\ref{lim}) can be rewritten using a bounded function
\begin{equation}\label{mu1}
\mu(z,w,\lambda):= \psi(z,w,\lambda)e^{-\lambda (z-w)^2/4},
\end{equation}
i.e., (\ref{fir})-(\ref{lim}) is equivalent to
\begin{equation}\label{fir2}
\overline{\partial }\mu(z,w,\lambda) = Q \overline{\mu}e^{[\overline{\lambda(z-w)^2}- \lambda(z-w)^2]/4}, \quad z\in \mathbb C; \quad \mu\rightarrow I, ~ z \rightarrow \infty.
\end{equation}
	Using  the fact that $\overline{\partial}\frac{1}{\pi z}=\delta(0)$, equation (\ref{fir2}) can be reduced to the Lippmann-Schwinger equation
\begin{equation}\label{19JanA}
\mu(z,\lambda,w)=I+ \frac{1}{\pi}\int_{\mathbb C}  Q(z') \frac{e^{-i\Im [\lambda(z'-w)^2]/2}}{z-z'}\overline{\mu}(z',\lambda,w) \, d{\sigma_{z'}},
\end{equation}
where $d{\sigma_{z'}}=dx'dy'$ and $\mu\to I$ as $z\to\infty$.

Denote
\begin{equation}\label{muSet5}
\mathcal L_\lambda \varphi (z)= \frac{1}{\pi}\int_{\mathbb C} \frac{e^{-i\Im[\lambda(z'-w)^2]/2} }{z-z'} \, \varphi(z') \, d{\sigma_{z'}}.
\end{equation}
Then equation (\ref{19JanA}) implies that
\begin{equation}\label{2006A}
\mu = I + \mathcal L_\lambda Q (I + \overline{\mathcal L_\lambda} \overline{Q} \mu ).
\end{equation}
In particular, for the component $\mu_{11}$ of the matrix $\mu$, we have
$ \mu_{11} = 1 + M \mu_{11}$, with
$ M = \mathcal L_\lambda Q_{12} \overline{\mathcal L_\lambda} \overline{Q_{21}},$
leading to
\begin{equation}\label{2410C}
(I-M)(\mu_{11} - 1) = M 1.
\end{equation}
By inverting $I-M$, we can obtain $\mu_{11}$. Other components of $\mu$ can be found similarly.

Denote by $L^\infty_{z,w}(B)$ the space of bounded functions of $z,w\in \mathbb C$ with values in a Banach space $B$.
The following two lemmas show that $M$ is a contractive operator in the space
$L^\infty_{z,w}(L^p_\lambda(\lambda:|\lambda|>R))$ if $R$ is large enough,
and that $M 1$ also belongs to this space. After these lemmas are proved, one can find the solution $\mu$ of (\ref{19JanA}) (using, for example, the Neumann series for the inversion of  $I-M$). Then formula (\ref{mu1}) provides the solution $\psi$ of (\ref{fir})-(\ref{lim}).
\begin{lemma}\label{2310A}
Let $p>2$. Then
$$ \lim_{R \rightarrow \infty} \|M\|_{L^\infty_{z,w}(L^p_\lambda(\lambda:|\lambda|>R)) } = 0. $$
\end{lemma}
\begin{lemma}\label{2310C}
Let $p>2$. Then there exists $R>0$ such that
$$ M 1 \in {L^\infty_{z,w}(L^p_\lambda(\lambda:|\lambda|>R))}. $$
\end{lemma}
Note that (\ref{2410C}) together with Lemmas \ref{2310A} and \ref{2310C} allows one to solve the direct but not the inverse problem, since operator $M$ depends on $Q$. The following inclusion is an immediate consequence of (\ref{2410C}) and Lemmas \ref{2310A} and \ref{2310C}:
\begin{equation}\label{mm1}
\mu_{11}-1 \in {L^\infty_{z,w}(L^p_\lambda(\lambda:|\lambda|>R))}, \quad p>2,
\end{equation}
for large enough $R$.

\subsection{Determination of the potential}

Let the matrix $h$ be the {\it (generalized) scattering data}, given by the formula
\begin{equation}\label{14Abr1}
{h}(\lambda,w)  =  \int_{\mathbb{C}}  e^{{-i\Im[\lambda(z-w)^2]/2}} Q(z)\overline{\mu}(z,\lambda,w) \, d{\sigma_{z}}.
\end{equation}
One can use Green's formula
 $$
\int_{\partial \mathcal O} f \, dz = 2i \int_{\mathcal O} \overline \partial {f} \, d{\sigma_{z}}
$$
to rewrite $h$ as
\begin{equation}\label{2106A}
  h(\lambda,w)  =\frac{1}{2i} \int_{\partial\mathcal O}{\mu}(z,\lambda,w) \, dz.
\end{equation}
Thus, one does not need to know the potential $Q$ in order to find $h$. Function $h$ can be evaluated if the Dirichlet data $\psi|_{\partial\mathcal O}$ is known for equation (\ref{fir}), since $\mu|_{\partial\mathcal O}$ in (\ref{2106A}) can be expressed via $\psi|_{\partial\mathcal O}$ using (\ref{mu1}).

The spectral parameter $i\lambda$ with real $\lambda$ was used in the standard approach to recover the potential from scattering data (\ref{14Abr1}), and the potential was
recovered by the limit of the scattering data as $\lambda\to\infty$. Instead, in the present work, we have $\lambda\in \mathbb C$, and
the potential is determined by integrating
the scattering data over a large annulus in the complex $\lambda$-plane.

Let $T^\lambda$  be the operator defined by
\begin{equation}\label{0911A}
T^\lambda [G]= \int_{\mathcal O} e^{-i \Im [\lambda(z-w)^2]/2} Q(z) G(z) \, d{\sigma_{z}},
\end{equation}
where $G$ can be a matrix- or scalar-valued function. Then
\begin{equation}\label{hhh}
 h(\lambda,w)=T^\lambda [\mu]= T^\lambda [I]+T^\lambda [\mu-I].
 \end{equation}
We will show that the following statement is valid.
\begin{theorem}\label{t23}
Let $Q$ be a complex-valued bounded potential. Then
\begin{equation}
\sup_{w\in \mathcal O}|\int_{R<|\lambda|<2R} |\lambda|^{-1} \, T^{\lambda} [\mu-I] \, d\sigma_\lambda| \rightarrow 0, \quad \text{as } \ R \to \infty, \label{remainder}
\end{equation}
and
\begin{equation}
\int_{\mathcal{O}} g(w)  \int_{R<|\lambda|<2R} |\lambda|^{-1} T^{\lambda} [I] \, d\sigma_\lambda \, d{\sigma_{w}} \to 4\pi^2 \ln 2\int_{\mathcal{O}} g(z) Q(z) \, d{\sigma_{z}}, \quad \text{as } R \to \infty, \label{mainTerm}
\end{equation}
for every smooth $g$ with a compact support in $\mathcal O$. Thus
\[
\int_{\mathcal{O}} g(z) Q(z)d\sigma_{z}=\frac{1}{4\pi^2\ln 2}\lim_{R\to\infty}\int_{R<|\lambda|<2R} |\lambda|^{-1} \, \int_{\mathbb{C}} g(w) h(\lambda,w)\, d{\sigma_{w}} d\sigma_\lambda.
\]

\end{theorem}
 Therefore, if the scattering data is uniquely
determined by the DtN map, then so is the potential $Q$.

In order to prove \eqref{remainder}, we use the two lemmas stated below
and (\ref{2410C}) rewritten as follows
\begin{align}\label{rem}
\mu_{11} - 1 = M (\mu_{11} - 1) + M 1
\end{align}
 (other entries of the matrix $\mu-I$ can be handled in a similar way). Relation \eqref{mainTerm} follows from the stationary phase approximation.

\begin{lemma}\label{2410D}
Let $p>1$. Then there exists $R>0$ such that
$$T^\lambda M 1 \in L^\infty_{w}(L^p_\lambda(\lambda:|\lambda|>R)).$$
\end{lemma}

\begin{lemma}\label{2410I}
Let $p>1$. Then there exists $R>0$ such that
$$T^\lambda M (\mu_{11} - 1) \in L^\infty_{w}(L^p_\lambda(\lambda:|\lambda|>R)).$$
\end{lemma}

\section{Proofs}\label{proofs}
In order to make the calculations more compact, we introduce the following notation for the $L^p$-space on the complement of the ball:
$$\quad L^p_{|\lambda|>R} = L^p_\lambda(\lambda:|\lambda|>R).$$
We will also use the real-valued function
$$\rho_{\lambda,w}(z) = \Im[\lambda(z-w)^2]/2,$$
where the dependence on $\lambda$ and $w$ will be omitted in some cases.
\subsection{Preliminary results}

\begin{lemma}\label{2210A}
Let $1\leq p<2$. Then the following estimate is valid for an arbitrary $0 \neq a \in \mathbb C$ and some constants $C=C(p,R)$ and $\delta=\delta(p)>0$:
$$
\left \| \frac{1}{u(\sqrt{u}-a)} \right \|_{L^p(u\in \mathbb C:|u|<R)} \leq C(1+|a|^{-1+\delta}).
$$
\end{lemma}
{\bf Remark.} A more accurate estimate will be proved below with $\delta=\frac{4}{p}-2$ if $1\leq p<4/3$, and with the right-hand side replaced by $C(1+|\ln|a||^{1/p})$ when $p=4/3$, or by a constant when $4/3<p<2$.

{\bf Proof.} The statement is obvious if $|a|\geq 1$. If $|a|<1$, then the left-hand side $L$ in the inequality above takes the following form after the substitution $u=|a|^2v$:
\begin{equation}\label{lll}
L=|a|^{\frac{4}{p}-3}\left \| \frac{1}{v(\sqrt{v}-\dot{a})} \right \|_{L^p(v\in \mathbb C:|v|<R/|a|^2)}, \quad \dot{a}=a/|a|.
\end{equation}

Without loss of the generality, one can assume that $R>2$. We split the function $f:=\frac{1}{v(\sqrt{v}-\dot{a})}$ into two terms $f_1+f_2$ obtained by multiplying $f$ by $\alpha$ and $1-\alpha$, respectively, where $\alpha$ is the indicator function of the disk of radius two. The norm of $f_1$ can be estimated from above by an $a$-independent constant. The second function can be estimated from above by $\frac{2}{|v|^{3/2}}$.  The norm of the latter function can be easily evaluated, and it does not exceed a constant if $p>4/3$. It does not exceed $C(1+|\ln|a||^{1/p})$ if $p=4/3$, and it does not exceed $C|a|^{3-\frac{4}{p}}$ if $p<4/3$. Since $\|f_1\|\leq C\|f_2\|$, we can replace $f$ in (\ref{lll}) by $Cf_2$, and this implies the statement of the lemma.
\qed

\begin{lemma}\label{2310F}
Let $z_1,w \in \mathbb C$, $p > 2$ and $\varphi \in L^\infty_{\rm{comp}}$. Then
$$
\left \|\int_{\mathbb C} \varphi(z)\frac{e^{i\rho_{\lambda,w}(z)} }{z-z_1} \, d{\sigma_{z}} \right \|_{L^p_\lambda(\mathbb C)}
\leq C\frac{\|\varphi\|_{L^\infty}}{|z_1-w|^{1-\delta}},
$$
where constant $C$ depends only on the support of $\varphi$ and on $\delta =\delta(p)>0$.
\end{lemma}
{\bf Proof}.
Denote by $F=F(\lambda,w,z_1)$ the integral in the left-hand side of the inequality above. We change variables $u=(z-w)^2$ in $F$ and take
into account that $d{\sigma_{u}} =4|z-w|^2 d{\sigma_{z}}$. Then
$$
F=\frac{1}{4}\sum_{\pm} \int_{\mathbb C} \varphi(w\pm\sqrt{u})\frac{e^{i\Im(\lambda u)/2}}{|u|(\pm\sqrt{u}-(z_1-w))} \, d{\sigma_{u}}.
$$
Using the Hausdorff-Young inequality with $p'=p/(p-1)$ and Lemma \ref{2210A}, we obtain that
$$
\|F\|_{L^p_\lambda} \leq \frac{1}{2}\sum_{\pm}\left \| \frac{\varphi(w\pm\sqrt{u})}{|u|(\pm\sqrt{u}-(z_1-w))} \right \|_{L^{p'}_u} \leq C \frac{\|\varphi\|_{L^\infty} }{|z_1-w|^{1-\delta}}.
$$
\qed

\subsection{Proof of Lemma \ref{2310A}}
Let
\begin{equation}\label{AAA}
A(z,z_2,\lambda,w) =\pi^{-2}\int_{\mathcal O}
 \frac{e^{-i\rho_{\lambda,w}(z_1)}}{{z}-{z_1}} {Q}_{12}(z_1) \frac{e^{i\rho_{\lambda,w}(z_2)}}{\overline{z_1}-\overline{z_2}} \overline{Q}_{21}(z_2) \, d{\sigma_{z_1}},
\end{equation}
so that
$$Mg(z) = \int_{\mathcal{O}} A(z,z_2,\lambda,w) g(z_2) \, d{\sigma_{z_2}}.$$
Then, from the Minkowski's integral inequality, we have
\begin{align*}
\|Mg(z,\cdot)\|_{L^p_{|\lambda|>R}}
&\leq \int_{\mathcal O} \|A(z,z_2,\lambda,w)g(z_2,\cdot)\|_{L^p_{|\lambda|>R}} \, d{\sigma_{z_2}} \\
&\leq \int_{\mathcal O} \sup_{\lambda:|\lambda|>R}|A(z,z_2,\lambda,w)| \, d{\sigma_{z_2}} \, \sup_{z_2}\|g(z_2,\cdot)\|_{L^p_{|\lambda|>R}}.
\end{align*}
Thus it remains to show that, uniformly in $z \in \mathbb C$ and $w \in \mathcal{O}$, we have
\[
\int_\mathcal O|A(z,z_2,\lambda,w)| \, d{\sigma_{z_2}} \to 0 \quad {\rm as} \quad |\lambda|\to\infty.
\]

Let $A^{s}$ be given by (\ref{AAA}) with the extra factor $\alpha(s|z-z_1|)\alpha(s|z_1-z_2|))$ in the integrand, where $\alpha\in C^\infty,~ \alpha=1$ outside of a neighborhood of the origin, and $\alpha$ vanishes in a smaller neighborhood of the origin.  Since
\begin{align*}
\int_{B_1(0)} \int_{B_1(0)} \frac{1}{|z_1|}\frac{1}{|z_1-z_2|} \, d{\sigma_{z_1}} \, d{\sigma_{z_2}} < \infty,
\end{align*}
for each $\varepsilon$ there exists $s=s_0(\varepsilon)$ such that
\[
\int_\mathcal O|A-A^{s_0}|\, d{\sigma_{z_2}}<\varepsilon
\]
for all the values of $z,w,\lambda$. Denote by $A^{s_0,n}$ the function $A^{s_0}$ with potentials ${Q}_{12},{Q}_{2 1}$ replaced by their $L_1$-approximations ${Q}_{12}^n,{Q}_{2 1}^n\in C_0^\infty$. Since the other factors in the integrand of  $A^{s_0}$ are bounded (they are infinitely smooth), we can choose these approximations in such a way that
\[
\int_\mathcal O|A^{s_0}-A^{s_0,n}|\, d{\sigma_{z_2}}<\varepsilon
\]
for all the values of $z,w,\lambda$. Now it is enough to show that
\[
|A^{s_0,n}(z,z_2,\lambda,w)|\to 0 \quad {\rm as} \quad |\lambda|\to \infty
\]
uniformly in $z,z_2,w$. The latter relation follows immediately from the stationary phase method, since the amplitude function in the integral $A^{s_0,n}$ and all the derivatives in $z_1$ of the amplitude function are uniformly bounded with respect to all the arguments.
\qed

\subsection{Proof of Lemma \ref{2310C}}
Recall that
\begin{align*}
M1 =\pi^{-2} \int_{\mathcal{O}} \int_{\mathcal{O}} \frac{e^{-i \rho_\lambda (z_1)}}{z-z_1} Q_{12}(z_1) \frac{e^{i \rho_\lambda (z_2)}}{\overline{z_1} - \overline{z_2}} \overline{Q_{21}}(z_2) \, d{\sigma_{z_2}} \, d{\sigma_{z_1}}.
\end{align*}
Let $C$ be a constant that may depend on $\| Q \|_{L^\infty}$ and $\mathcal{O}$.
Then, by Minkowski's integral inequality	and Lemma \ref{2310F}, we have
\begin{eqnarray*}
\|M 1\|_{L^p_{|\lambda|>R}}
&\leq&  \int_{\mathcal O} \left \| \frac{e^{-i \rho_\lambda (z_1)}}{z-z_1} Q_{12}(z_1) \int_{\mathcal O}
\frac{e^{i\rho_\lambda(z_2)}}{\overline{z_1}-\overline{z_2}}\overline{Q_{21}}(z_2) \, d{\sigma_{z_2}} \right \|_{L^p_{|\lambda|>R}} d{\sigma_{z_1}} \\
&\leq&  \int_{\mathcal O} \left | \frac{Q_{12}(z_1)}{z-z_1}  \right |\left \|\int_{\mathcal O}
\frac{e^{i\rho_\lambda(z_2)}}{\overline{z_1}-\overline{z_2}}\overline{Q_{21}}(z_2) \, d{\sigma_{z_2}} \right \|_{L^p_{|\lambda|>R}} d{\sigma_{z_1}} \\
&\leq& \nonumber C\int_{\mathcal O}
  \frac{1}{|z-z_1||z_1-w|^{1-\delta}} \, d{\sigma_{z_1}}  < \infty,
\end{eqnarray*}
since $\delta>0$.
\qed

\subsection{Proof of Lemma \ref{2410D}}
Let $C$ be a constant that may depend on $\| Q \|_{L^\infty}$ and $\mathcal{O}$. Then, applying successively Minkowski's integral inequality, Holder's inequality, and Lemma \ref{2310F}, we see that
\begin{align*}
\|T^\lambda[M1]\|_{L^p_{|\lambda|>R}} & \leq \int_{\mathcal O} \left \|\int_{\mathcal O}
\frac{e^{-i (\rho (z_1) + \rho(z))}}{z-z_1} Q(z) \, d{\sigma_{z}} \int_{\mathcal O}\frac{e^{i\rho (z_2)}}{\overline{z_1}-\overline{z_2}}
\overline{Q_{21}}(z_2) \, d{\sigma_{z_2}} \right \|_{L^p_{|\lambda|>R}}|Q_{12}(z_1)| d{\sigma_{z_1}} \\
& \leq C\int_{\mathcal O} \left \|\int_{\mathcal O}
 \frac{e^{-i \rho (z)}}{z-z_1} Q(z) \, d{\sigma_{z}} \right \|_{L^{2p}_{|\lambda|>R}} \left \| \int_{\mathcal O}\frac{e^{i\rho (z_2)}}{\overline{z_1}-\overline{z_2}}
\overline{Q_{21}}(z_2) \, d{\sigma_{z_2}} \right \|_{L^{2p}_{|\lambda|>R}} d{\sigma_{z_1}} \\
& \leq C\int_{\mathcal O} \frac{1}{|z_1-w|^{1-\delta}}
\frac{1}{|z_1-w|^{1-\delta}} \, d{\sigma_{z_1}}  < \infty,
\end{align*}
as $\delta>0$.
\qed

\subsection{Proof of Lemma \ref{2410I}}
Let $f = \mu_{11}-1$ and let $C$ be a constant that may depend on $\| Q \|_{L^\infty}$ and $\mathcal{O}$. Then the same arguments as in the proof of Lemma \ref{2410D} imply that
\begin{align*}
\|T^\lambda[M f]\|_{L^p_{|\lambda|>R}} & \leq
C\int_{\mathcal O} \left \|\int_{\mathcal O}
 \frac{e^{-i \rho(z)}}{z-z_1} Q(z) d{\sigma_{z}} \right \|_{L^{2p}_{|\lambda|>R}} \left \| \int_{\mathcal O}\frac{e^{i\rho(z_2)}}{\overline{z_1}-\overline{z_2}}
\overline{Q_{21}}(z_2) f(z_2) d{\sigma_{z_2}} \right \|_{L^{2p}_{|\lambda|>R}} d{\sigma_{z_1}} \\
& \leq C \int_{\mathcal O} \left \|\int_{\mathcal O}
 \frac{e^{-i \rho(z)}}{z-z_1} Q(z) d{\sigma_{z}} \right \|_{L^{2p}_{|\lambda|>R}} \int_{\mathcal O} \left |\frac{\overline{Q_{21}}(z_2) }{\overline{z_1}-\overline{z_2}}
\right | \left \| f(z_2) \right \|_{L^{2p}_{|\lambda|>R}} d{\sigma_{z_2}} d{\sigma_{z_1}} \\
&  \leq C\|f\|_{L^\infty_{z,w}\left(L^{2p}_{|\lambda|>R} \right)}\int_{\mathcal O} \frac{1}{|z_1-w|^{1-\delta}} \, d{\sigma_{z_1}} < \infty,
\end{align*}
since $\delta>0$ and (\ref{mm1}) holds for $f=\mu_{11}-1$.
\qed

\subsection{Proof of Theorem \ref{t23}}
Let us prove (\ref{remainder}). We fix $p\in(1,2)$. From (\ref{rem}) and Lemmas \ref{2410D} and \ref{2410I}, it follows that
 there exists $R>0$ such that
$ T^\lambda[\mu_{11}-1]\in L^\infty_{w}(L^p_{|\lambda|>R})$. Other entries of matrix $\mu-I$ can be treated similarly, i.e.,
\[
T^\lambda[\mu-I]\in L^\infty_{w}(L^p_{|\lambda|>R}).
\]
Since $q=\frac{p}{p-1}>2$, Holder's inequality implies that
\[
|\int_{R<|\lambda|<2R} |\lambda|^{-1} \, T^{\lambda} [\mu-I] \, d\sigma_\lambda|\leq[\int_{R<|\lambda|<2R} |\lambda|^{-q}d\sigma_\lambda]^{\frac{1}{q}} \|T^{\lambda} [ \mu-I]\|_{L^\infty_{w}(L^p_{|\lambda|>R})}\to 0
\]
as $R\to\infty.$ Relation  (\ref{remainder}) is proved.

The stationary phase approximation implies that
\[
\int_{\mathcal O}T^\lambda[1]g(w)d\sigma_w=\int_{\mathcal O} \int_{\mathcal O}e^{-i \Im [\lambda(z-w)^2]/2} g(w)d\sigma_w Q(z) \, d{\sigma_{z}}=\int_{\mathcal O}[\frac{2\pi}{|\lambda|}g(z)+O(|\lambda|^\frac{-3}{2})]Q(z)d\sigma_z.
\]
This immediately justifies (\ref{mainTerm}). The last statement of the theorem follows from (\ref{hhh})-(\ref{mainTerm}).
\qed

\subsection{Proof of Theorem \ref{uniqueness}}

Due to Theorem \ref{t23}, one only needs to show that the scattering data $h$ for $|\lambda|\gg 1 $ is uniquely determined by the Dirichlet-to-Neumann operator $\Lambda_{\gamma}$. This will be done by repeating the arguments used in \cite[Theorem 4.1]{bu} and \cite[Theorem 5.1]{fr}.

Let $\gamma_j, j=1,2,$ be two Lipshitz conductivities in $\mathcal O$ such that $\Lambda_{\gamma_1} = \Lambda_{\gamma_2}$.
Since $\gamma_j$ is Lipschitz continuous, it is differentiable almost everywhere, and the derivatives are bounded \cite{ev}.
Since $\Lambda_{\gamma_1} = \Lambda_{\gamma_2}$ and $\gamma_1, \gamma_2 \in W^{1,\infty}(\mathcal{O})$, we have $\gamma_1 |_{\partial \mathcal{O}} = \gamma_2 |_{\partial \mathcal{O}}$ (see \cite{a90}).
We extend $\gamma_j$ outside $\mathcal{O}$ in such a way that $\gamma_1=\gamma_2$ in $\mathbb{C} \setminus \mathcal{O}$ and $1-\gamma_j \in W^{1, \infty}_{comp}(\mathbb{C})$. Let $\widetilde{\mathcal O}$ be a bounded domain with a smooth boundary that contains supports of functions $1-\gamma_j$. All the previous results will be used below with $\mathcal{O}$ replaced by $\widetilde{\mathcal{O}}$ and $\gamma$ extended as described above. Let $Q_j, \psi_j, \mu_j, h_j,~j=1,2,$ be the potential and the solution in \eqref{fir}, the function in \eqref{mu1}, and the scattering data in \eqref{14Abr1} associated with the extended conductivity $\gamma_j$. Let us note that functions
$\psi_j, \mu_j, h_j,~j=1,2,$ defined by the conductivity problem in $\widetilde{\mathcal{O}}$ are not extensions of the functions defined by the problem in $\mathcal{O}$.

Due to equation \eqref{2106A}, we have
\begin{align*}
  h_j(\lambda,w)  =\frac{1}{2i} \int_{\partial\widetilde{\mathcal O}}{\mu_j}(z,\lambda,w) \, dz.
\end{align*}
Thus it is enough to prove that
\begin{align}\label{mumu}
\mu_1 = \mu_2 \quad \text{on } \partial \widetilde{\mathcal{O}}\quad \text{when } |\lambda|\gg 1 .
\end{align}

 Let $\varphi = (\varphi_1, \varphi_2)^t$ be the first column of $\psi_1$ and $v = \gamma_1^{-1/2} \varphi_1$, $w = \gamma_1^{-1/2} \overline{\varphi_2}$. Since  $\overline{\partial} \varphi = Q_1 \overline{\varphi} $, and equation (\ref{firbc}) holds for $\phi^{(1)} = (\varphi_1, \overline{\varphi_2})^t$, it follows that
$\overline{\partial} v = \partial w$ in $\mathbb C$,
and therefore there exists $u_1$ such that
\begin{align*}
\partial u_1 = v,  \quad \overline{\partial} u_1 = w \quad \text{in } \mathbb C,
\end{align*}
which is a solution to
\begin{align*}
\mbox{div}(\gamma_1 \nabla u_1) =0 \quad \text{in } \mathbb C.
\end{align*}
Now we define $u_2$ by
\begin{align*}
u_2= \begin{cases}
u_1 \quad \text{in } \mathbb{C} \setminus \mathcal{O} \\
\widehat{u} \quad \text{in } \mathcal{O},
\end{cases}
\end{align*}
where $\widehat{u}$ is the solution to the Dirichlet problem
\begin{align*}
\begin{cases}
\mbox{div}(\gamma_2 \nabla \widehat{u}) =0 & \text{in } \mathcal{O} \\
\widehat{u} = u_1 & \text{on } \partial \mathcal{O}.
\end{cases}
\end{align*}
Let $g \in C^\infty_0 (\mathbb{C})$. Then
\begin{align*}
\int_\mathbb{C} \gamma_2 \nabla u_2 \nabla g \, d\sigma_z &=
\int_{\mathbb C \setminus \mathcal{O}} \gamma_1 \nabla u_1 \nabla g \, d\sigma_z + \int_\mathcal{O} \gamma_2 \nabla \widehat{u} \nabla g \, d\sigma_z \\
&=- \int_{\partial \mathcal{O}} \Lambda_{\gamma_1} [u_1 |_{\partial \mathcal{O}}] g \, dz + \int_{\partial \mathcal{O}} \Lambda_{\gamma_2} [\widehat{u} |_{\partial \mathcal{O}}] g \, dz \\
&= 0.
\end{align*}
Hence $\mbox{div}(\gamma_2 \nabla u_2) =0 $ in $ \mathbb{C}$. Then
\begin{align*}
\phi^{(2)} = \gamma_2^{1/2} \left( \partial u_2, \overline{\partial }u_2 \right)^t
\end{align*}
is the solution of (\ref{firbc}) with $\gamma=\gamma_2$, and
\begin{align*}
\varphi^{(2)} = (\phi^{(2)},\overline{\phi^{(2)}})^t
\end{align*}
is the solution of (\ref{fir}) with $Q=Q_2$.

Lemmas \ref{2310A} and \ref{2310C} imply the unique solvability of the Lippmann-Schwinger equation when $|\lambda|>R$ and $R$ is large enough. Thus, $\varphi^{(2)}$ is equal to the first column of $\psi_2$ when $|\lambda|>R$. On the other hand, $\varphi^{(2)}$  in $\mathbb C\setminus\mathcal{O}$ coincides with the first column $\varphi$ of $\psi_1$. Thus  the first columns of $\psi_1$ and $\psi_2$ are equal on $\mathbb C\setminus\mathcal{O}$ when $|\lambda|>R$. Repeating the same steps with the second columns of $\psi_1,\psi_2$, we obtain that $\psi_1|_{\partial\widetilde{\mathcal{O}}}=\psi_2|_{\partial\widetilde{\mathcal{O}}}$ when $|\lambda|>R$, and therefore (\ref{mumu}) holds.

The uniqueness of $h$ and Theorem \ref{t23} imply that the potential $Q$ in the Dirac equation (\ref{fir}) is defined uniquely, and therefore $q$ is  defined uniquely. Now the conductivity $\gamma$ can be found from (\ref{char1bc}) uniquely up to an additive constant. Finally, this constant can be defined uniquely since $\gamma |_{\partial \mathcal{O}}$ is defined uniquely by $\Lambda_\gamma$.

\qed

{\bf Acknowledgments.}  The authors are thankful to Daniel Faraco and Keith Rogers for useful discussions.

\end{document}